\documentclass[showpacs,10pt,twocolumn,prb]{revtex4-1}
\usepackage{amsmath}
\usepackage{amssymb}
\usepackage{graphics}
\usepackage{epsfig}

\begin{document}

\title{Spin glass behavior of semiconducting K$_{x}$Fe$_{2-y}$S$_{2}$}
\author{Hechang Lei, Milinda Abeykoon, Emil S. Bozin, and C. Petrovic}
\affiliation{Condensed Matter Physics and Materials Science Department, Brookhaven
National Laboratory, Upton, NY 11973, USA}
\date{\today}

\begin{abstract}
We report discovery of K$_{x}$Fe$_{2-y}$S$_{2}$ single crystals,
isostructural to K$_{x}$Fe$_{2-y}$Se$_{2}$ superconductors. The sulfide
compound is a small gap semiconductor and shows spin glass behavior below 32
K. Our results indicate that stoichiometry, defects and local environment of
FeCh (Ch = S, Se) tetrahedra have important effects on the physical
properties of isostructural and isoelectronic K$_{x}$Fe$_{2-y}$Ch$_{2}$
compounds.
\end{abstract}

\maketitle

Iron based materials are in the focus of exploratory search for new
superconductors since the discovery of LaFeAsO$_{1-x}$F$_{x}$ with
transition temperature $T_{c}$ up to 26 K.\cite{Kamihara} Several
superconducting families were discovered soon after REFePnO (RE = rare
earth; Pn = P or As, FePn-1111 type),\cite{Rotter}$^{-}$\cite{Hsu FC}
including $\alpha $-PbO type FeCh (Ch = S, Se, and Te, FeCh-11 type)
materials that do not have any crystallographic layer in between puckered
Fe-Ch slabs.\cite{Hsu FC} FeCh-11 type materials share a square-planar
lattice of Fe with tetrahedral coordination and similar Fermi surface
topology with other iron-based superconductors.\cite{Subedi} Under external
pressure,\cite{Mizuguchi2}$^{,}$\cite{Medvedev} the $T_{c}$ can be increased
from 8 K to 37 K and the $dT_{c}/dP$ can reach 9.1 K/GPa, the highest in all
iron-based superconductors.\cite{Medvedev} The empirical rule proposed by
Mizuguchi et al. proposes that the critical temperature is closely
correlated with the anion height between Fe and Ch layers. There is an
optimal distance around 0.138 nm with a maximum transition temperature $%
T_{c}\simeq $55 K.\cite{Mizuguchi3}

Intercalation can change the local environment of Fe-Se tetrahedron and
introduce extra carriers. The intercalation could also decrease
dimensionality of conducting bands. This is favorable for superconductivity
since the presence of low energy electronic collective modes in layered
conductors helps to screen Coulomb interaction.\cite{Bill} This is seen in
iron based superconductors: the $T_{c}$ increases from FeCh-11 type to
FePn-1111 type. Very recently the superconducting $T_{c}$\ is enhanced in
iron selenide material to about 30 K not by external pressure but by
inserting K, Rb, Cs, and Tl between the FeSe layers (AFeSe-122 type), thus
changing the crystal structure around FeCh tetrahedra.\cite{Guo}$^{-}$\cite%
{Fang MH} Similar to pressure effects, the intercalation using elements with
+1 valence decreases Se height towards the optimum value.\cite%
{Krzton-Maziopa} The expanded Fe-Se interlayer distances could also
contribute to reducing dimensionality of conducting bands and magnetic
interactions. On the other hand, the insulating-superconducting transition
(IST) can be induced in (Tl$_{1-x}$K$_{x}$)Fe$_{2-y}$Se$_{2}$ by tuning the
Fe stoichiometry and implying that the superconductivity is in proximity of
an antiferromagnetic (AFM) Mott insulating state.\cite{Fang MH} Thus
exploring new oxychalcogenide and chalcogenide compounds containing similar
FeCh layers would be instructive.

In this work, we report discovery of K$_{x}$Fe$_{2-y}$S$_{2}$ single
crystals isostructural to 122 iron selenide superconductors. The structure
analysis indicates that the anion height might not be essential for
superconductivity. The resistivity and magnetic measurements suggest spin
glass (SG) semiconductor ground state similar to the TlFe$_{2-x}$Se$_{2}$
with high Fe deficiency, even though anion height are close to values found
in iron based superconductors with $T_{c}$ above 20 K.\cite{Fang MH}$^{,}$\cite{Ying JJ2}

Single crystals of K$_{x}$Fe$_{2-y}$S$_{2}$ were grown by self-flux method%
\cite{Kihou} with nominal composition K:Fe:S = 0.8:2:2. Prereacted FeS and K
pieces were added into the alumina crucible with partial pressure of argon
gas. The quartz tubes were heated to 1030 ${^{\circ }}C$, kept at this
temperature for 3 hours, then cooled to 730 ${^{\circ }}C$. Platelike
crystals up to 10$\times $10$\times $3 mm$^{3}$ can be grown. Powder X-ray
diffraction (XRD) data were collected at 300 K using 0.3184 \AA\ wavelength
radiation (38.94 keV) at X7B beamline of the National Synchrotron Light
Source. The average stoichiometry was determined by energy-dispersive x-ray
spectroscopy (EDX). Electrical transport, heat capacity, and magnetization
measurements were carried out in Quantum Design PPMS-9 and MPMS-XL5.

Fig. 1(a) shows powder XRD data at the room temperature and structural
refinements on K$_{x}$Fe$_{2-y}$S$_{2}$ using General Structure Analysis System (GSAS).\cite{Larson}$^{,}$\cite%
{Toby} Model possessing tetragonal ThCr$_{2}$Si$_{2}$ structure, space group
I4/mmm, failed to explain the observed diffraction pattern, due to clear
appearance of (110) and other superlattice reflections indicating symmetry
lowering to I4/m. Data were successfully explained within I4/m symmetry that
incorporates Fe vacancy order site, with lattice parameters a = 8.3984(5)
\AA\ and c = 13.5988(11) \AA , appreciably smaller than those observed in
the selenium counterpart.\cite{Guo} The c axis is particularly reduced due
to the smaller ionic size of S$^{2-}$ when compared to Se$^{2-}$. The
ordered Fe vacancy is the same as in K$_{x}$Fe$_{2-y}$Se$_{2}$.\cite{Bao W}
This may imply similar origin of magnetic behavior for both compounds.
Atomic positions with refined parameters are listed in Table 1. Refinements
yielded that K1, K2 and Fe1 positions are partially occupied, while Fe2 are
almost fully occupied. It should be noted that when K1 and Fe1 positions are
fully unoccupied while K2 and Fe2 are fully occupied, the corresponding
chemical formula is K$_{0.8}$Fe$_{1.6}$S$_{2}$ and the Fe vacancy is
completely ordered. The average atomic ratios from EDX are consistent with K$%
_{0.88(6)}$Fe$_{1.63(4)}$S$_{2.00(1)}$, indicating that there are both
potassium and iron deficiencies from ideal 122 stoichiometry, in good
agreement with XRD fitting results. It should be noted that the 18.5\% of
the iron precipitates on the surface of the ingot on cooling, The iron
precipitates are easy to remove and have no influence on the physical
properties of K$_{x}$Fe$_{2-y}$S$_{2}$.

\begin{figure}[tbp]
\centerline{\includegraphics[scale=1.0]{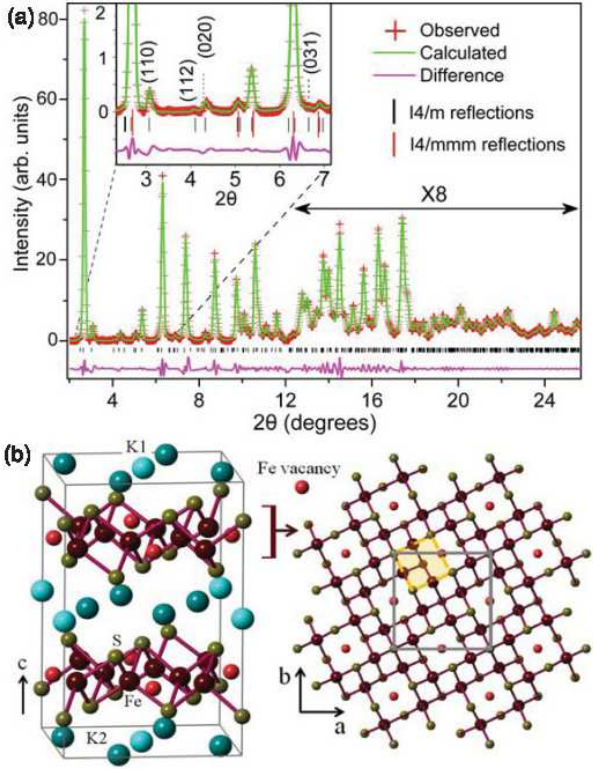}} \vspace*{-0.3cm}
\caption{(a) Powder XRD patterns of K$_{x}$Fe$_{2-y}$S$_{2}$ and fit using
I4/m model. Inset: low scattering angle part, emphasizing presence of
superlattice reflections characteristic of I4/m symmetry. Enlarged isotropic
thermal parameters, particularly in the potassium layer, are indicative of
local disorder being present in the structure. (b) Crystal structure of K$%
_{x}$Fe$_{2-y}$S$_{2}$ in I4/m unit cell with vacant Fe1 sites marked red
and K1 sites marked light blue (left). Sketch of the FeS slab (c-axis view),
with ordered Fe vacancies (right). Yellow and grey squares illustrate I4/mmm
and I4/m unit cells respectively. }
\end{figure}

\begin{table}[tbp]\centering%
\caption{Structural parameters for K$_{x}$Fe$_{2-y}$S$_{2}$ at room
temperature. Values in brackets give the number of equivalent distances or
angles of each type.}%
\begin{tabular}{cccccc}
\hline\hline
\multicolumn{3}{c}{Chemical Formula} & \multicolumn{3}{c}{K$_{0.88}$Fe$%
_{1.63}$S$_{2}$} \\
\multicolumn{3}{c}{Space Group} & \multicolumn{3}{c}{I4/m} \\
\multicolumn{3}{c}{a (\AA )} & \multicolumn{3}{c}{8.3984(5)} \\
\multicolumn{3}{c}{c (\AA )} & \multicolumn{3}{c}{13.5988(11)} \\
\multicolumn{3}{c}{V (\AA $^{3}$)} & \multicolumn{3}{c}{959.17(11)} \\ \hline
\multicolumn{3}{c}{Interatomic Distances (\AA )} & \multicolumn{3}{c}{Bond
Angles ($^{\circ }$)} \\
\multicolumn{2}{c}{d$_{Fe1-S2}$ [4]} & 2.4170(12) & \multicolumn{2}{c}{
S2-Fe1-S2 [2]} & 110.1(5) \\
\multicolumn{2}{c}{d$_{Fe1-Fe2}$ [4]} & 2.5914(16) & \multicolumn{2}{c}{
S2-Fe1-S2 [4]} & 109.2(3) \\
\multicolumn{2}{c}{d$_{Fe2-S1}$ [1]} & 2.3647(11) & \multicolumn{2}{c}{
S1-Fe2-S2 [1]} & 103.9(5) \\
\multicolumn{2}{c}{d$_{Fe2-S2}$ [1]} & 2.3369(11) & \multicolumn{2}{c}{
S1-Fe2-S2 [1]} & 109.8(3) \\
\multicolumn{2}{c}{d$_{Fe2-S2}$ [1]} & 2.3005(11) & \multicolumn{2}{c}{
S1-Fe2-S2 [1]} & 111.0(3) \\
\multicolumn{2}{c}{d$_{Fe2-S2}$ [1]} & 2.2660(11) & \multicolumn{2}{c}{
S2-Fe2-S2 [1]} & 105.3(3) \\
\multicolumn{2}{c}{d$_{Fe2-Fe2}$ [2]} & 2.6495(16) & \multicolumn{2}{c}{
S2-Fe2-S2 [1]} & 117.7(2) \\
\multicolumn{2}{c}{d$_{Fe2-Fe2}$ [1]} & 2.8135(17) & \multicolumn{2}{c}{
S2-Fe2-S2 [1]} & 108.7(5) \\ \hline
\multicolumn{6}{c}{Anion Heights (\AA )} \\
\multicolumn{2}{c}{S1 to Fe1} & 1.388(5) & \multicolumn{2}{c}{S2 to Fe1} &
1.384(5) \\
\multicolumn{2}{c}{S1 to Fe2} & 1.334(5) & \multicolumn{2}{c}{S2 to Fe2} &
1.439(5) \\ \hline
Atom & x & y & z & Occ & U$_{iso}$ (\AA $^{2}$) \\
K1 & 0 & 0 & 0.5 & 0.84(12) & 0.059(15) \\
K2 & 0.80(2) & 0.418(6) & 0.5000 & 0.89(3) & 0.059(15) \\
Fe1 & 0 & 0.5 & 0.25 & 0.08(4) & 0.0136(3) \\
Fe2 & 0.2954(5) & 0.4111(5) & 0.2460(16) & 1.00(1) & 0.0136(3) \\
S1 & 0 & 0 & 0.1479(5) & 1.00(0) & 0.0111(6) \\
S2 & 0.1113(4) & 0.292(1) & 0.3518(3) & 1.00(0) & 0.0111(6) \\ \hline\hline
\end{tabular}%
\label{TableKey}%
\end{table}%

The in-plane resistivity $\rho _{ab}$(T) of the K$_{x}$Fe$_{2-y}$S$_{2}$
single crystal rapidly increases with decreasing the temperature from $\rho
_{ab}$(300 K) $\sim $ 100 m$\Omega $ cm and there is no obvious
magnetoresistance (Fig. 2). The $\rho _{ab}$(T) is thermally activated: $%
\rho =\rho _{0}\exp (E_{a}/k_{B}T)$, where $\rho _{0}$ is a prefactor and $%
k_{B}$ is the Boltzmann's constant (inset (b) of Fig 2). Using the $\rho
_{ab}$(T) data from 70 to 300 K, we estimate $\rho _{0}$ $\sim $ 11.7(2) m$%
\Omega $ cm and the activation energy $E_{a}$ = 51.8(2) meV. The
semiconducting behavior might at least partially be ascribed to the
deficiency of Fe in Fe-Se plane that would introduce random scattering
potential, just like in highly Fe deficient K$_{x}$Fe$_{2-y}$Se$_{2}$ and
TlFe$_{2-x}$Se$_{2}$.\cite{Fang MH}$^{,}$\cite{Ying JJ2}$^{,}$\cite{Wang DM}

\begin{figure}[tbp]
\centerline{\includegraphics[scale=0.7]{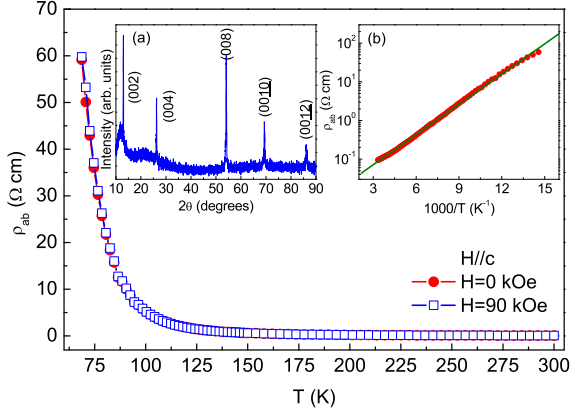}} \vspace*{-0.3cm}
\caption{Temperature dependence of the in-plane resistivity $\protect\rho %
_{ab}(T)$ with H = 0 (closed red circles) and 90 kOe (open blue squares).
Inset (a) shows the single crystal XRD pattern of K$_{x}$Fe$_{2-y}$S$_{2}$
obtained using Rigaku Miniflex. The crystal surface is normal to the c axis
with the plate-shaped surface parallel to the ab-plane. Inset (b) shows the
fitting result using thermally activated model for $\protect\rho _{ab}(T)$
in zero field.}
\end{figure}

The magnetic susceptibility with H$\Vert $ab is larger than with H$\Vert $c (Fig. 3(a)), similar
to observed anisotropy in TlFe$_{2-x}$Se$_{2}$.\cite{Ying JJ2} The most
interesting characteristics are the absence of Curie-Weiss behavior and
obvious bifurcation between the zero-field-cooling (ZFC) and field-cooling
(FC) curves below 32 K. This might suggest the presence of low-dimensional
short range magnetic correlations and/or a long range magnetic order above
300 K, and an antiferromagnetic phase transition at low temperatures. The
M(T) irreversible behavior below 32 K implies some ferromagnetic
contribution to magnetic susceptibility or a glassy transition where spins
would be frozen randomly below the freezing temperature $T_{f}$. Similar
magnetization has been reported in\ TlFe$_{2-x}$Se$_{2}$ and KFeCuS$_{2}$.%
\cite{Ying JJ2}$^{,}$\cite{Oledzka} Inset in Fig. 3 (a) shows the
magnetization loops for H$\Vert $c. At 250 K, the M-H loop is almost linear
and there is no hysteresis. However, an s-shape M-H loop can be observed at
1.8 K, which is a typical behavior of a SG system.\cite{Ying JJ2} The
s-shape M-H loop is present at T = 50 K which indicates that short-range
ferromagnetic interaction may exist above $T_{f}$. As shown in Fig. 3(b),
the peak in the real part of ac susceptibility $\chi ^{\prime }(T)$ exhibits
strong frequency dependence in ac magnetic field. When the frequency
increases, the peak positions shift to higher temperatures whereas the
magnitudes decrease, indicating a typical SG behavior.\cite{Mydosh} By
fitting the frequency dependence of the peak shift using $K=\Delta
T_{f}/(T_{f}\Delta \log f)$, we obtained K = 0.0134(5). This is in agreement
with values (0.0045 $\leqslant $ K $\leqslant $ 0.08) found in the canonical
SG system, but much smaller than in typical superparamagnet.\cite{Mydosh}
Fig. 3(c) shows the magnetic field dependence of the thermoremanent
magnetization (TRM). The sample was cooled from T = 60 K (above $T_{f}$) in
a magnetic field to T = 10 K (below $T_{f}$) and then kept at 10 K for a $%
t_{w}$ = 100 s. Then, the magnetic field was removed and the magnetization
decay $M_{TRM}(t)$ was measured. It can be seen that, below $T_{f}$ (T = 10
K), $M_{TRM}(t)$ decays slowly so that its value is non-zero even after
several hours. This is another signature of the SG behavior, i.e., the
existence of extremely slow spin relaxation below $T_{f}$.\cite{Mydosh} In
contrast, above $T_{f}$, $M_{TRM}(t)$ quickly relaxes and does not show slow
decay (inset (a) of Fig. 3(c)). The magnetization decay can be explained
well using a stretched exponential function commonly used to explain TRM
behavior in SG systems, $M_{TRM}(t)=M_{0}\exp [-(t/\tau )^{1-n}]$, where $%
M_{0}$, $\tau $, and $1-n$ are the glassy component, the relaxation
characteristic time, and the critical exponent, respectively. It can be seen
(inset (b) of Fig. 3(c)) that $\tau $ decreases significantly with field but
the 1-n increases slightly. On the other hand, the value of 1-n is close to
1/3, consistent with theoretical predictions and the experiments on
traditional SG system.\cite{Campbell}$^{,}$\cite{Chu} The SG behavior could
originate from Fe clusters induced by vacancies and disorder, and the
exchange interactions between spins within a cluster would depend on the
distribution of iron ions (Table 1).\cite{Oledzka} Indeed, in TlFe$_{2-x}$Se$%
_{2}$, the ground state is a reentrant spin glass if the content of Fe is
below 1.7.\cite{Ying JJ2} However, for x values larger than 1.7, TlFe$_{2-x}$%
Se$_{2}$ becomes a superconductor below 20 K.\cite{Fang MH} Therefore
superconductivity in K$_{x}$Fe$_{2-y}$S$_{2}$ might be induced for smaller
deficiency of Fe.

\begin{figure}[tbp]
\centerline{\includegraphics[scale=0.7]{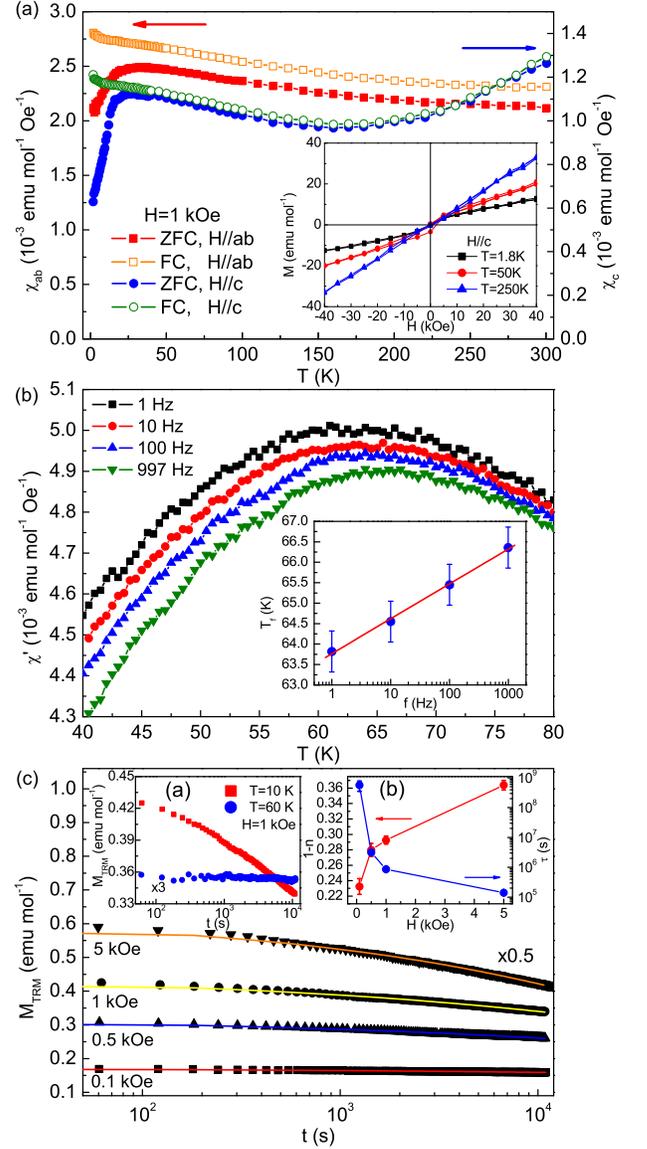}} \vspace*{-0.3cm}
\caption{(a) ZFC and FC dc magnetic susceptibility with H$\Vert $c and H$%
\Vert $ab below 300 K. Inset: isothermal M(H) for H$\Vert $c at T = 1.8, 50,
and 250 K. (b) Temperature dependence of $\protect\chi ^{\prime }(T)$
measured at several fixed frequencies. Inset: the frequency dependence of $%
T_{f}$. The solid line is the linear fit to the $T_{f}$ data. (c) $M_{TRM}$
vs. $t$ at 10 K with various dc fields and $t_{w}$ = 100 s. The solid lines
are fits using a stretched exponential function. Inset (a): $M_{TRM}$ vs. $t$
at 10 K and 60 K with H = 1 kOe and $t_{w}$ = 100 s. Inset (b): magnetic
field dependence of $1-n$ and $\protect\tau $.}
\end{figure}

Specific heat of K$_{x}$Fe$_{2-y}$S$_{2}$ (Fig. 4) approaches the
Dulong-Petit value of 3NR at high temperature, where N is the atomic number
in the chemical formula (N = 5) and R is the gas constant. At \ low
temperature, specific heat can be fitted using $C_{p}$ = $\gamma
_{SG}T+\beta T^{3}$ (inset (a) of Fig. 4). The $\gamma _{SG}$ is commonly
found in magnetic insulating SG system, implying constant density of states
of the low temperature magnetic excitations.\cite{Huang}$^{-}$\cite{Raju}
The second term is due to phonon contribution. The obtained $\gamma _{SG}$\
is 1.58(6) mJ/mol-K$^{2}$. The Debye temperatures $\Theta _{D}$ can be
calculated from $\beta $ through $\Theta _{D}$ = $(12\pi ^{4}NR/5\beta
)^{1/3}$ to be $\Theta _{D}$ = 284.3(7) K. It should be noted that, as
opposed to usual $\lambda $ anomaly, there is a very weak broad hump of C$%
_{p}$/T near T = 32.6 K (inset (b) of Fig. 4). This is expected for bulk low
dimensional or glassy magnetic systems.\cite{Huang}$^{,}$\cite{Brodale}$^{,}$%
\cite{Martin}

\begin{figure}[tbp]
\centerline{\includegraphics[scale=0.9]{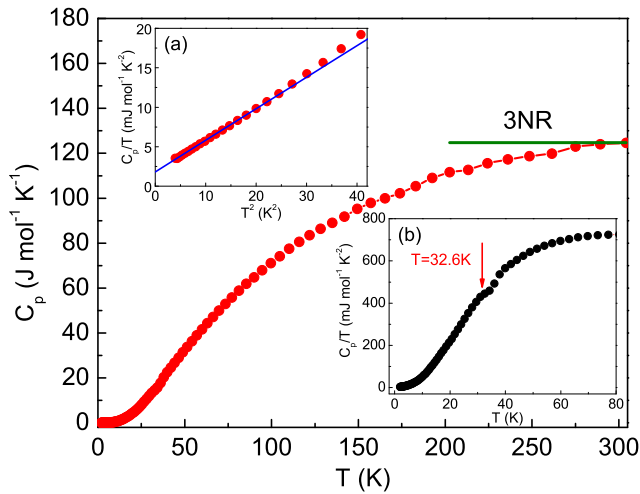}} \vspace*{-0.3cm}
\caption{Temperature dependence of specific heat. Inset (a) shows the
low-temperature specific-heat data in the plot of C$_{p}$/T vs T$^{2}$. The
blue solid line is the fitting curves using formula C$_{p}$/T = $\protect%
\gamma_{SG} $+$\protect\beta $T$^{2}$. Inset (b) shows the enlarged area
near magnetic transition region of C$_{p}$/T-T.}
\end{figure}

Discovery of K$_{x}$Fe$_{2-y}$S$_{2}$ implies that it is possible to tune
conductivity and magnetism by changing chalcogen elements (S and Se) in
isostructural K$_{x}$Fe$_{2-y}$Ch$_{2}$ materials. According to the Table 1,
anion heights S1 to Fe1 and S1 to Fe2 in K$_{0.88(6)}$Fe$_{1.63(4)}$S$%
_{2.00(1)}$ are close to the alleged optimal value of 1.38 \r{A}, comparable
to distances in FePn-1111 type materials and lower than in K$_{x}$Fe$_{2-y}$%
Se$_{2}$.\cite{Mizuguchi3}$^{,}$\cite{Krzton-Maziopa} Hence, if the anion
height is the crucial parameter, the sulfide compound should be a
superconductor with higher $T_{c}$ than K$_{x}$Fe$_{2-y}$Se$_{2}$. However,
it is a semiconductor not a superconductor.\ We also note that in K$_{x}$Fe$%
_{2-y}$S$_{2}$ there are two Fe sites and two corresponding anion heights.
Our results indicate that the anion height values may be not essential
parameters that govern superconductivity in AFeSe-122 compounds. In
contrast, it suggests that disorder and site occupancies are significantly
important.

In summary, we report discovery of K$_{x}$Fe$_{2-y}$S$_{2}$ single crystals
isostructural to T$_{c}$ = 33 K superconductor K$_{x}$Fe$_{2-y}$Se$_{2}$ and
exhibiting similar width of formation and vacancies on both potassium and
iron site. Sulfide material is semiconducting and glassy magnetic suggesting
that the physical properties are governed by stoichiometry, defects, local
environment of Fe-S tetrahedra.

We thank S. L. Bud'ko for discussions, John Warren for help with SEM
measurements and John Hanson for help in facilitating X7B experiment. Work
at Brookhaven is supported by the U.S. DOE under Contract No.
DE-AC02-98CH10886 and in part by the Center for Emergent Superconductivity,
an Energy Frontier Research Center funded by the U.S. DOE, Office for Basic
Energy Science.


\begin{thebibliography}{99}
\bibitem{Kamihara} Y. Kamihara, T. Watanabe, M. Hirano, and H. Hosono, J.
Am. Chem. Soc. \textbf{130}, 3296 (2008).

\bibitem{Rotter} M. Rotter, M. Tegel, and D. Johrendt, Phys. Rev. Lett.
\textbf{101}, 107006 (2008).

\bibitem{Wang XC} X. C. Wang, Q. Q. Liu, Y. X. Lv, W. B. Gao, L. X. Yang, R.
C. Yu, F. Y. Li, and C. Q. Jin, Solid State Commun. \textbf{148}, 538 (2008).

\bibitem{Hsu FC} F. C. Hsu, J. Y. Luo, K. W. Yeh, T. K. Chen, T. W. Huang,
P. M. Wu, Y. C. Lee, Y. L. Huang, Y. Y. Chu, D. C. Yan, and M. K. Wu, Proc.
Natl. Acad. Sci. USA \textbf{105}, 14262 (2008).

\bibitem{Subedi} A. Subedi, L. Zhang, D. J. Singh, and M. H. Du, Phys. Rev.
B \textbf{78}, 134514 (2008) .

\bibitem{Mizuguchi2} Y. Mizuguchi, F. Tomioka, S. Tsuda, T. Yamaguchi, and
Y. Takano, Appl. Phys. Lett. \textbf{93}, 152505 (2008).

\bibitem{Medvedev} S. Medvedev, T. M. McQueen, I. Trojan, T. Palasyuk, M. I.
Eremets, R. J. Cava, S. Naghavi, F. Casper, V. Ksenofontov, G. Wortmann, and
C. Felser, Nature Mater. \textbf{8,} 630 (2009).

\bibitem{Mizuguchi3} Y. Mizuguchi, Y. Hara, K. Deguchi, S. Tsuda, T.
Yamaguchi, K. Takeda, H. Kotegawa, H. Tou, and Y. Takano, Supercond. Sci.
Technol. \textbf{23,} 054013 (2010).

\bibitem{Bill} A. Bill, \ H. Morawitz, and V. Z. Kresin, Phys. Rev. B
\textbf{68}, 144519 (2003)

\bibitem{Guo} J. Guo, S. Jin, G. Wang, S. Wang, K. Zhu, T. Zhou, M. He, and
X. Chen, Phys. Rev. B \textbf{82}, 180520(R) (2010).

\bibitem{Wang AF} A. F. Wang, J. J. Ying, Y. J. Yan, R. H. Liu, X. G. Luo,
Z. Y. Li, X. F. Wang, M. Zhang, G. J. Ye, P. Cheng, Z. J. Xiang, and X. H.
Chen, Phys. Rev. B \textbf{83}, 060512(R) (2011).

\bibitem{Krzton-Maziopa} A. Krzton-Maziopa, Z. Shermadini, E. Pomjakushina,
V. Pomjakushin, M. Bendele, A. Amato, R. Khasanov, H. Luetkens, and K.
Conder, J. Phys.: Condens. Matter \textbf{23}, 052203 (2011).

\bibitem{Fang MH} M. H. Fang, H. D. Wang, C. H. Dong, Z. J. Li, C. M. Feng,
J. Chen, and H. Q. Yuan, Europhys. Lett. \textbf{94}, 27009 (2011).

\bibitem{Ying JJ2} J. J. Ying, A. F. Wang, Z. J. Xiang, X. G. Luo, R. H.
Liu, X. F. Wang, Y. J. Yan, M. Zhang, G. J. Ye, P. Cheng, and X. H. Chen,
arXiv:1012.2929.

\bibitem{Kihou} K. Kihou, T. Saito, S. Ishida, M. Nakajima, Y. Tomioka, H.
Fukazawa, Y. Kohori, T. Ito, S. Uchida, A. Iyo, C. Lee, and H. Eisaki, J.
Phys. Soc. Jpn. \textbf{79}, 124713 (2010).

\bibitem{Larson} A. C. Larson and R. B. Von Dreele, Los Alamos National
Laboratory Report LAUR 86-748 (1994).

\bibitem{Toby} B. H. Toby, J. Appl. Cryst. \textbf{34}, 210 (2001).

\bibitem{Bao W} W. Bao, Q. Huang, G. F. Chen, M. A. Green, D. M. Wang, J. B.
He, X. Q. Wang, and Y. Qiu, arXiv:1102.0830 (2011).

\bibitem{Wang DM} D. M. Wang, J. B. He, T.-L. Xia, and G. F. Chen, Phys. Rev. B 83, 132502 (2011).

\bibitem{Oledzka} M. Oledzka, K. V. Ramanujachary, and M. Greenblatt, Mater.
Res. Bull. \textbf{31,} 1491 (1996).

\bibitem{Mydosh} J. A. Mydosh, Spin Glasses: An Experimental Introduction
Taylor \& Francis, London, 1993.

\bibitem{Campbell} I. A. Campbell, Phys. Rev. B \textbf{37}, 9800 (1988).

\bibitem{Chu} D. Chu, G. G. Kenning, and R. Orbach, Phys. Rev. Lett. \textbf{%
72}, 3270 (1994).

\bibitem{Huang} C. Y. Huang, J. Magn. Magn. Mater. \textbf{51}, 1 (1985).

\bibitem{Meschede} D. Meschede, F. Steglich, W. Felsch, H. Maletta, and W.
Zinn, Phys. Rev. Lett. \textbf{44}, 102 (1980).

\bibitem{Raju} N. P. Raju, E. Gmelin, and R. K. Kremer, Phys. Rev. B \textbf{%
46}, 5405 (1992).

\bibitem{Brodale} G. E. Brodale, R. A. Fisher, W. E. Fogle, N. E. Phillips,
and J. van Curen, J. Magn. Magn. Mater. \textbf{31-34}, 1331 (1983).

\bibitem{Martin} L. O.-S. Martin, J. P. Chapman, L. Lezama, J. J. S.
Garitaonandia, J. S. Marcos, J. Rodr\'{\i}guez-Fern\'{a}ndez, M. I.
Arriortua, and T. Rojo, J. Mater. Chem. \textbf{16}, 66 (2006).
\end{thebibliography}
\end{document}